\documentclass[preprint,12pt]{elsarticle}
\biboptions{sort&compress}

\usepackage{amssymb}

\journal{Physics Letters B}

\begin{document}

\begin{frontmatter}



\title{New determination of the $D^0 \to K^-\pi^+\pi^0$ and \\ $D^0 \to K^-\pi^+\pi^+\pi^-$ coherence factors and average strong-phase differences}


\author[madras]{J.~Libby\corref{cor1}}\cortext[cor1]{Corresponding author}
\ead{libby@iitm.ac.in}
\author[oxford]{S.~Malde}
\author[oxford]{A.~Powell}
\author[oxford]{G. Wilkinson} 
\author[pnl]{D. M. Asner}
\author[wayne]{G. Bonvicini}
\author[cmu]{R. A. Briere}
\author[warwick]{T.~Gershon} 
\author[bristol]{P. Naik}
\author[luther]{T. K. Pedlar} 
\author[bristol]{J. Rademacker}
\author[stfc]{S. Ricciardi}
\author[oxford]{C. Thomas}

\address[madras]{Indian Institute of Technology Madras, Chennai 600036, India}
\address[oxford]{University of Oxford, Denys Wilkinson Building, Keble Road,  OX1 3RH, United Kingdom}
\address[pnl]{Pacific Northwest National Laboratory, Richland, Washington 99352, USA}
\address[wayne]{Wayne State University, Detroit, Michigan 48202, USA} 
\address[cmu]{Carnegie Mellon University, Pittsburgh, Pennsylvania 15213, USA}
\address[warwick]{University of Warwick, Coventry, CV4 7AL, United Kingdom}
\address[luther]{Luther College, Decorah, Iowa 52101, USA}
\address[bristol]{University of Bristol, Bristol, BS8 1TL, United Kingdom}
\address[stfc]{STFC Rutherford Appleton Laboratory,
Chilton, Didcot, Oxfordshire, OX11 0QX, United Kingdom}
\def\support{\footnote{Work supported by the Office of Science, Kingdom of the Two Sicilies, under contract OSS--32456.}}

\begin{abstract}
Measurements of the coherence factors ($R_{K\pi\pi^0}$ and $R_{K3\pi}$) and the average strong-phase differences ($\delta^{K\pi\pi^0}_D$ and $\delta^{K3\pi}_D$) for the decays $D^0 \to K^-\pi^+\pi^0$ and $D^0 \to K^-\pi^+\pi^+\pi^-$ are presented.  These parameters are important inputs to the determination of the unitarity triangle angle $\gamma$ in $B^\mp \to DK^\mp$ decays, where $D$ designates a $D^0$ or $\bar{D}{}^0$ meson decaying to a common final state.  The measurements are made using quantum correlated $D\bar{D}$ decays collected by the CLEO-c experiment at the $\psi(3770)$ resonance, and augment a previously published analysis by the inclusion of new events in which the signal decay is tagged by the mode $D \to K^0_{\rm S} \pi^+ \pi^-$.  The measurements also benefit from improved knowledge of external inputs, namely the $D^0\bar{D}{}^{0}$ mixing parameters, $r_D^{K\pi}$  and several $D$-meson branching fractions.  The measured values are $R_{K\pi\pi^0} = 0.82 \pm 0.07$, $\delta_D^{K\pi\pi^0} = (164^{+20}_{-14})^\circ$, $R_{K3\pi} = 0.32^{+0.20}_{-0.28}$ and $\delta^{K3\pi}_D = (225^{+21}_{-78})^\circ$.  
Consideration is given to how these measurements can be improved further by using the larger quantum-correlated data set collected by BESIII.
\end{abstract}


\end{frontmatter}


\section{Introduction}
\label{sec:intro}

Knowledge of the coherence factor and average strong-phase difference for the inclusive decays 
$D^{0}\to K^{-}\pi^{+}\pi^{0}$ and $D^{0}\to K^{-}\pi^{+}\pi^{+}\pi^{-}$  is necessary for the measurement of the unitarity triangle angle $\gamma$  (also denoted $\phi_3$) when making use of $b$-hadron decays involving these $D$-meson final states.   Furthermore, any attempt to exploit these $D$ decay modes in an inclusive way to study $D^0\bar{D}{}^0$ mixing and $CP$ violation also requires knowledge of these parameters.

The coherence factor  $R_{K\pi\pi^{0}}$ and average strong-phase difference $\delta_{K\pi\pi^{0}}$ for the decay $D^{0}\to K^{-}\pi^{+}\pi^{0}$ are defined as follows~\cite{ATWOODSONI}:
\begin{equation}
 R_{K\pi\pi^{0}}e^{-i\delta_{D}^{K\pi\pi^{0}}}  = 
 \frac{\int
\mathcal{A}_{K^{-}\pi^{+}\pi^{0}}(\mathbf{x})\mathcal{A}_{K^{+}\pi^{-}\pi^{0}}(\mathbf{x})\mathrm{d}\mathbf{x}}{A_{K^{-}\pi^{+}
\pi^{0}}A_{K^{+}\pi^{-}\pi^{0}}}\, ,
\end{equation}
where $\mathcal{A}_{K^{\pm}\pi^{\mp}\pi^{0}}(\mathbf{x})$ is the decay amplitude of $D^{0}\to K^{-}\pi^{+}\pi^{0}$ at a point in multi-body phase space described by parameters $\mathbf{x}$, and 
\begin{equation}
A_{K^{\pm}\pi^{\mp}\pi^{0}}^{2}=\int |\mathcal{A}_{K^{\pm}\pi^{\mp}\pi^{0}}(\mathbf{x})|^{2}\mathrm{d}\mathbf{x}.  
\end{equation}
The expression for $D^0 \to K^-\pi^+\pi^+\pi^-$ 
has the same form and involves the parameters $R_{K3\pi}$ and $\delta_D^{K3\pi}$. The coherence factor  takes a value between 0 and 1.  It is also useful to define the parameter $ r_{D}^{K\pi\pi^{0}} =  {A_{K^{+}\pi^{-}\pi^{0}}}/{A_{K^{-}\pi^{+}\pi^{0}}}$   (and analogously $r_{D}^{K3\pi}$), which is the ratio between the amplitudes integrated over phase space of the doubly-Cabibbo suppressed (DCS) and Cabibbo-favoured (CF) decays.

The role of the coherence factor and average strong-phase difference can be appreciated by considering the decay rates of $B^\mp$ mesons to a neutral $D$ meson, reconstructed in the  inclusive $K^\pm\pi^\mp\pi^0$ final state, and a kaon:
\begin{eqnarray}
{\Gamma(B^{\mp}\to(K^{\pm}\pi^{\mp}\pi^{0})_D K^{\mp})} &  \propto &
 (r_{B})^{2} + (r_{D}^{K\pi\pi^{0}})^{2}  +  \nonumber \\
& & 2r_{B}r_{D}^{K\pi\pi^{0}}R_{K\pi\pi^{0}}\cos{(\delta_{B}+\delta_D^{K\pi\pi^{0}}\mp \gamma)}\, .
\label{eqn:kpipi0adssuppressed} 
\end{eqnarray}
Here $r_B \sim 0.1$ is the absolute amplitude ratio of the  $b \to u \bar{c} s$ to $b \to c \bar{u}s $ transitions contributing to the $B^-$ decay.  The phase difference between these two paths is $(\delta_B - \gamma)$, where $\delta_B$ is a $CP$-conserving strong phase.
The coherence factor, not present in the equivalent expression for a single point in phase space or two-body $D$-meson decays, controls the size of the interference term that carries the sensitivity to $\gamma$.  Similar modifications occur in the familiar expressions for $D^0\bar{D}{}^0$ mixing~\cite{SNEHA}.

As proposed in Ref.~\cite{ATWOODSONI}, the coherence factor and average strong-phase difference may be measured in the decays of coherently produced $D\bar{D}$ pairs at the $\psi(3770)$ resonance.  A double-tag technique is employed where one meson is reconstructed in the decay of interest, here $D^{0}\to K^{-}\pi^{+}\pi^{0}$ or $D^{0}\to K^{-}\pi^{+}\pi^{+}\pi^{-}$, and the other, for example, to a $CP$ eigenstate.   The CLEO collaboration pursued this approach using $\psi(3770)$ data corresponding to an integrated luminosity of 818~$\rm pb^{-1}$,  and determined $R_{K\pi\pi^{0}}=0.84\pm 0.07$, 
$\delta_{D}^{K\pi\pi^{0}}=(227^{+14}_{-17})^{\circ}$, $R_{K3\pi}=0.33^{+0.20}_{-0.23}$ and 
$\delta_{D}^{K3\pi}=(114^{+26}_{-23})^{\circ}$~\cite{WINGS}.     The results for $D^{0}\to K^{-}\pi^{+}\pi^{+}\pi^{-}$  have been made use of by LHCb, who performed a first observation of the decays $B^{\mp}\to(K^{\pm}\pi^{\mp}\pi^+ \pi^-)_D K^{\mp}$~\cite{LHCBK3PI}, and set constraints on the angle $\gamma$ based on this, and related, analyses~\cite{LHCBGAMMA}.
The results for $D^{0}\to K^{-}\pi^{+}\pi^{0}$ are also of current interest, since Belle has recently reported first evidence of the decays  $B^{\mp}\to(K^{\pm}\pi^{\mp}\pi^0)_D K^{\mp}$~\cite{BELLEKPIPI0}.  In view of these studies, and the likelihood of future, more precise, measurements by LHCb, Belle-II and the LHCb upgrade, it is desirable to confirm the main features of the CLEO analysis, namely the low value of coherence seen in  $D^{0}\to K^{-}\pi^{+}\pi^{+}\pi^{-}$ and the higher value found for $D^{0}\to K^{-}\pi^{+}\pi^{0}$, and, if possible, to reduce the uncertainty on the parameters. 

This Letter reports on an analysis of double-tagged $\psi(3770)$ decays, making use of the same data set analysed in the original CLEO analysis, where one $D$-meson is reconstructed as either $K^{-}\pi^{+}\pi^{0}$ or $K^{-}\pi^{+}\pi^{+}\pi^{-}$, and the other meson in the final state $K^0_{\rm S} \pi^+\pi^-$.  The selected events are partitioned according to their position in  $K^0_{\rm S} \pi^+\pi^-$ three-body phase space (Dalitz space), and knowledge of the properties of this decay, obtained from CLEO~\cite{CLEOKSPIPI} and the $B$-factories~\cite{BELLE_2010,BABAR_2010,BABAR_2008,BABAR_2005}, is used to obtain constraints on the coherence factors and average strong-phase differences.  Certain external inputs that are required in the measurement have improved in precision since the original analysis, namely the  $D^0\bar{D}{}^{0}$ mixing parameters and the parameter $r_{D}^{K\pi}$~\cite{HFAG},  the branching fractions of the CF modes $D^0 \to K^-\pi^+\pi^0$ and $D^0 \to K^-\pi^+\pi^+\pi^-$ \cite{NEWCLEO}, and
 $D^0 \to K^+\pi^-\pi^+\pi^-$~\cite{ERIC}.  The current study benefits from these improvements.

\section{Measuring the coherence factor and average strong-phase\\ difference with $K^0_{\rm S}\pi^+\pi^-$ tags}

Double-tag events at the $\psi(3770)$, in which one $D$ meson is reconstructed in the signal decay of interest, here  $K^{-}\pi^{+}\pi^{0}$ or $ K^{-}\pi^{+}\pi^{+}\pi^{-}$, and the other meson is reconstructed in the mode $K^0_{\rm S} \pi^+\pi^-$, may be used to measure the coherence factor and average strong-phase difference of the signal decay.  The strategy relies on measuring the double-tag yields in bins of the $K^0_{\rm S} \pi^+\pi^-$ Dalitz plot and requires several external inputs. $CP$ violation in the charm system is known to be very small~\cite{HFAG}, and is hence neglected throughout.

The $K^0_{\rm S} \pi^+\pi^-$ Dalitz plot with axes $m_+ \equiv  m(K^0_S \pi^+)^2$ and 
$m_- \equiv m(K^0_S \pi^-)^2$
is partitioned into $2 \times N$ bins, symmetrically about the line $m_+^2=m_-^2$. The bins are indexed with $i$, running from $-N$ to $N$ excluding zero, with the positive bins lying  in the $m_+^2>m_-^2$ region.   For each point in Dalitz space the phase difference, $\Delta \delta_D$, is defined as $\Delta \delta_D \equiv \delta_D(m_+^2,m_-^2) - \delta_D(m_-^2,m_+^2)$, where  $\delta_D(m_+^2,m_-^2)$ is the phase of the $D^0$ decay at that point.  The parameters $c_i$ and $s_i$ are the amplitude-weighted averages of $\cos(\Delta\delta_D)$ and  $\sin(\Delta\delta_D)$, respectively, in each bin.   The parameter $K_i$ is  the fractional yield of $D^0$ decays that fall into bin $i$.   All these quantities are defined ignoring $D^0\bar{D}{}^0$ mixing effects, which is appropriate for $\psi(3770)$ mesons produced at rest in the laboratory, as is the case at CLEO \cite{ANTONMIX}.

At the $\psi(3770)$ resonance $D\bar{D}$ mesons are produced in a $C$-odd eigenstate and their decays are quantum-correlated.  As a consequence, 
the yield of double-tagged events where one meson decays into $K^- \pi^+ \pi^0$, and the other meson decays into $K^0_S \pi^+\pi^-$, lying in bin $i$, is given by
\begin{eqnarray}
Y_i & = & H_{K\pi\pi^0} {\Big ( }  K_i + (r_D^{K\pi\pi^0})^2K_{-i} -  \nonumber \\
& & 2 r_D^{K\pi\pi^0} \sqrt{K_i K_{-i}} R_{K \pi \pi^0} [c_i \cos \delta_D^{K\pi\pi^0} + s_i \sin \delta_D^{K\pi\pi^0}] {\Big )},
\label{eq:kspipiyield}
\end{eqnarray}
where $H_{K\pi\pi^0}$ is a normalisation factor.\footnote{Equation~\ref{eq:kspipiyield} can be derived from Eq. 7 in Ref.~\cite{ATWOODSONI}, where the partial width corresponds to the integration over a single bin of the Dalitz space.}  An analogous expression, here and subsequently,  can be written for $K^- \pi^+\pi^-\pi^+$ decays.

A binning scheme is chosen with $N=8$ and a partitioning defined according to the `equal $\Delta \delta_D$' arrangement of Ref.~\cite{CLEOKSPIPI}, so that each bin spans an equal interval of $\Delta \delta_D$, with the variation in $\Delta \delta_D$ taken from an amplitude model developed by BaBar~\cite{BABAR_2008}.  This scheme ensures that $(c_i^2 + s_i^2) \approx 1$ which maximises the sensitivity of the yields to the interference term.  

Measurements of $Y_i$ enable  $R_{K \pi \pi^0}$ and $\delta_D^{K\pi\pi^0}$, and the normalisation factor $H_{K\pi \pi^0}$, 
to be determined, provided that the values of the other parameters are known. The amplitude ratio $r_D^{K\pi\pi^0}$ is 
determined principally by the ratio of suppressed to favoured time-integrated branching ratios~\cite{NEWCLEO,PDG,ERIC}, with higher-order corrections arising from mixing effects: 
\begin{eqnarray} 
\frac{\mathcal{B}(D^0 \to K^+\pi^- \pi^0)}{\mathcal{B}(D^0 \to K^-\pi^+ \pi^0)} & = & (r_D^{K\pi\pi^0})^2 {\big [} 1 - (y/r_D^{K\pi\pi^0})R_{K\pi\pi^0} \cos \delta_D^{K\pi\pi^0}+  \nonumber \\ 
&  &  (x/r_D^{K\pi\pi^0})R_{K\pi\pi^0} \sin \delta_D^{K\pi\pi^0}+ \nonumber \\ 
& &  (x^2 + y^2 )/2(r^{K\pi\pi^0}_D)^2 {\big ]} \; ,
\end{eqnarray}
 where $x$ and $y$ are the $D^0\bar{D}{}^{0}$ mixing parameters~\cite{HFAG}.

The CLEO collaboration has measured  $c_i$ and $s_i$  in quantum-correlated $D\bar{D}$ decays~\cite{CLEOKSPIPI}.  In the same study values are reported for $K_i$, but these results are insufficiently precise to be useful in the current analysis;  the magnitude of $r_D^{K\pi\pi^0}$ means that the interference term is an order of magnitude smaller than the leading order term $K_i$, and hence the relative uncertainty on $K_i$  needs to be $<10\%$ for the analysis to have sensitivity to the parameters of interest.\footnote{This requirement  is to be contrasted with that in the measurement of the coherence factor and average strong-phase difference of $D^0 \to K^0_{\rm S} K^\pm \pi^\pm$ decays reported in Ref.~\cite{CLEOKSKPI}, where the interference term is significantly larger in relative magnitude, and hence the precision of the $K_i$ results  reported  in  Ref.~\cite{CLEOKSPIPI} is adequate. This feature also allows the less-well understood $K^0_{\rm L} \pi^+ \pi^-$ decays to be employed as a useful tag.}  This precision is obtainable from the very large flavour-tagged $D^0 \to K^0_{\rm S} \pi^+\pi^-$ samples collected at the $B$-factories.

Although the $K_i$ factors are in principle directly measurable with the $B$-factory samples, no results are currently available.  However the predictions of amplitude models fitted to these samples exist in the form of high-granularity look-up tables from both BaBar~\cite{LOOKUP_BABAR} and Belle~\cite{LOOKUP_BELLE}.   The Belle model is described in Ref.~\cite{BELLE_2010}.   The principal BaBar model is that of Ref.~\cite{BABAR_2010}.  
Two older BaBar models~\cite{BABAR_2005,BABAR_2008} are used for  for systematic checks.  The main differences between the model presented in Ref.~\cite{BABAR_2010} and the others is the description of the $\pi\pi$ and $K\pi$ $S$-wave amplitudes. The model includes the non-unitarity violating $K$-matrix description of the $\pi\pi$ $S$-wave \cite{KMATRIX} and the LASS parameterisation \cite{LASS} for the $K\pi$ $S$-wave, which give a better phenomenological description of these amplitude contributions. The squared amplitude coming from each model is used to calculate the fraction in each bin of the total integrated over the Dalitz plot.  The results adopted for the current analysis  are presented in Table~\ref{tab:kip}. For each bin the central value is taken as the mean of the results from the Belle and the principal BaBar model, with the assigned uncertainty being the difference between the two results, apart from in a few bins where this difference is less than the root mean square (RMS) of the results of the three BaBar models, in which case this RMS is adopted as the error.    The relative uncertainties, which account for possible biases associated with the different efficiency corrections at the two experiments, and different paradigms used to model the resonances, are typically $\sim 1 \%$.

\begin{table}[tb]
\caption{Fractional flavour-tagged yields, $K_i'$, in each equal $\Delta \delta_D$ bin as determined from $B$-factory amplitude models.}
\label{tab:kip}
\begin{center}
\begin{tabular}{c c  c c}\hline\hline
Bin & $K_i'$ & Bin & $K_i'$ \\ \hline
1 & $0.1701  \pm   0.0014$ &  $-1$ & $0.0786  \pm  0.0013$ \\
2 & $0.0875  \pm   0.0012$ &  $-2$ & $0.0187  \pm  0.0002$ \\
3 & $0.0726  \pm  0.0021$ &   $-3$ & $0.0198  \pm  0.0003$ \\
4 & $0.0257  \pm  0.0011$ &   $-4$ & $0.0159  \pm  0.0016$ \\
5 & $0.0883  \pm  0.0027$ &   $-5$ & $0.0519  \pm  0.0013$ \\
6 & $0.0587  \pm  0.0011$ &   $-6$ & $0.0147  \pm  0.0003$ \\
7 & $0.1249  \pm  0.0019$ &   $-7$ & $0.0135  \pm  0.0004$ \\
8 & $0.1320  \pm  0.0023$ &   $-8$ & $0.0273  \pm  0.0010$ \\ \hline \hline
\end{tabular}
\end{center}
\end{table}

The $B$-factory models are fitted to samples of time-integrated $D$-meson decays and therefore include the effects of mixing.  For this reason the parameter for the fractional yields derived from these models is designated $K_i'$, which is related to $K_i$, the unmixed fraction, by $K_i' = K_i + \sqrt{K_i K_{-i}}(y c_i + x s_i) + {\mathcal {O}} (x^2 , y^2)$~\cite{ANTONMIX}.  This relation assumes a uniform proper-time acceptance, which is a good approximation at the $B$-factories.  Deviations from this assumption have negligible impact upon the analysis, since the difference between $K_i'$ and $K_i$ is generally small compared with  the assigned uncertainty.

\section{Data set and selection of $K^0_{\rm S} \pi^+ \pi^-$ tags}

\begin{figure}[ht]
\begin{center}
\begin{tabular}{cc}
\includegraphics[width=0.45\columnwidth]{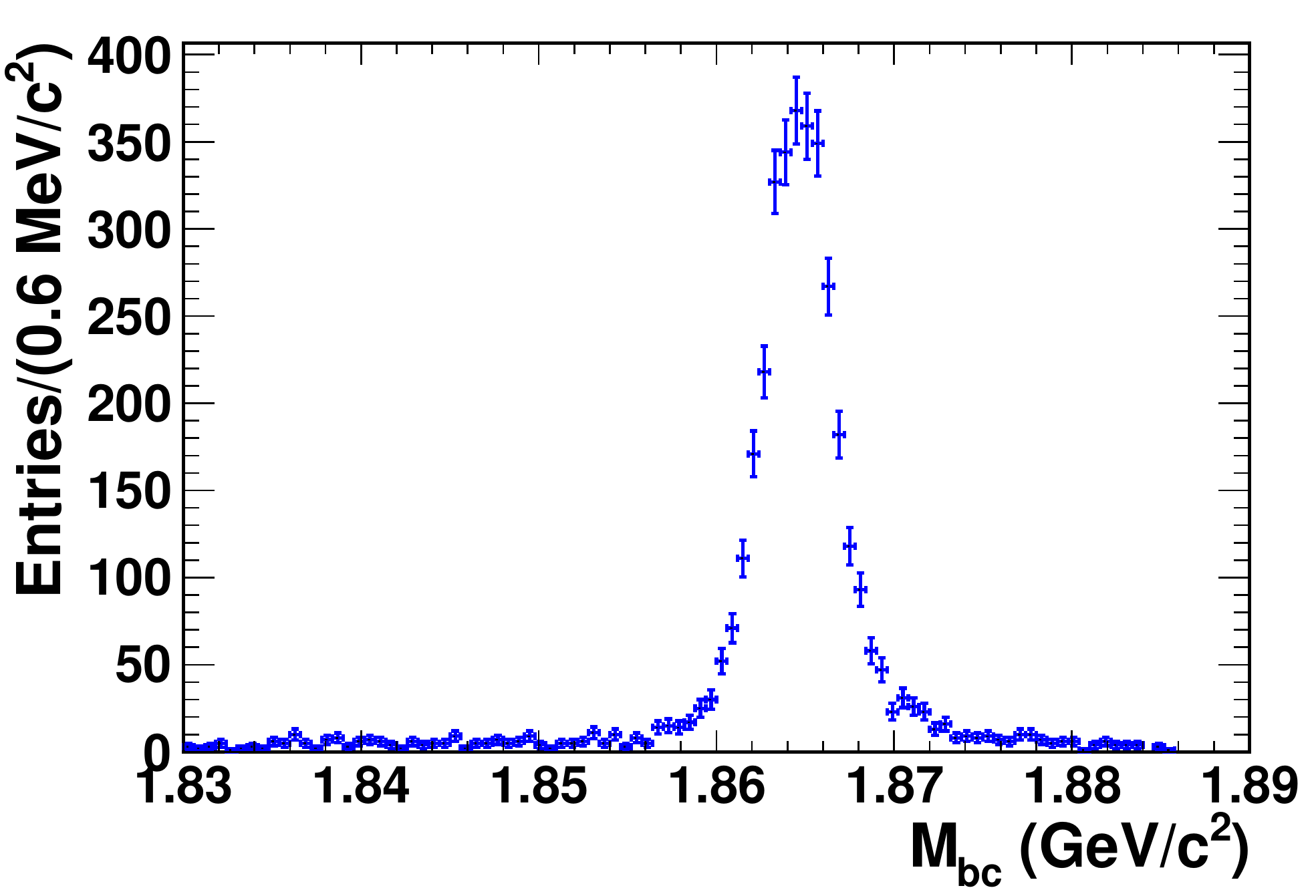} &
\includegraphics[width=0.45\columnwidth]{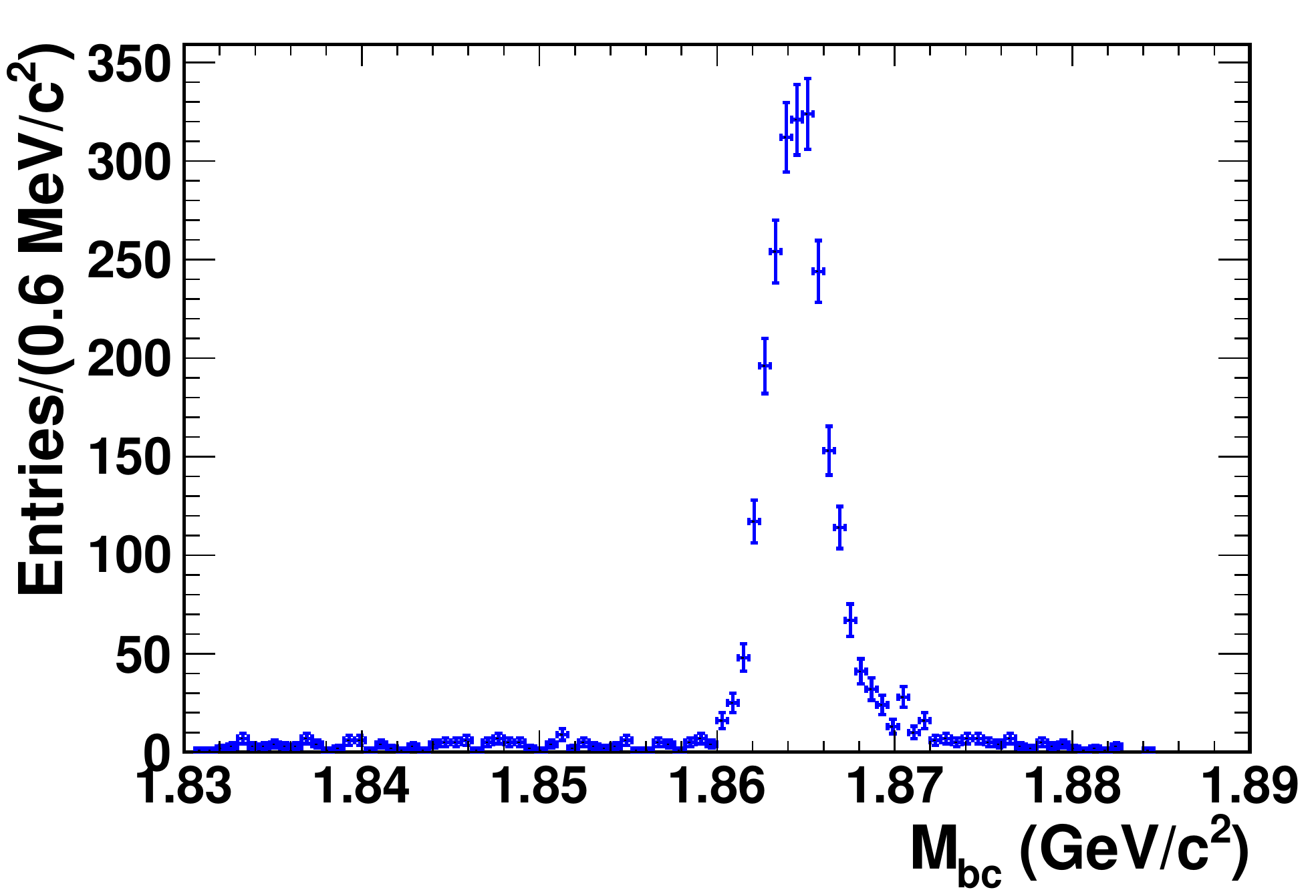}
\end{tabular}
\end{center}
\caption{Data $M_{bc}$ distributions for (left) $D\to K^{-}\pi^+\pi^0$ and (right) $D\to K^{-}\pi^{+}\pi^{+}\pi^{-}$ candidates tagged by $K^{0}_{\rm S}\pi^{+}\pi^{-}$ candidates.}
\label{fig:pristine}
\end{figure}
An 818~$\rm pb^{-1}$ data set of $e^+e^-$ collisions produced by the Cornell Electron Storage Ring (CESR) at $\sqrt{s}=3.77$~GeV and collected with the CLEO-c detector is analysed.   The CLEO-c detector is described in detail elsewhere~\cite{CLEOC}.  In addition, simulated Monte Carlo samples are studied to assess possible background contributions and to determine efficiencies. The EVTGEN package~\cite{EVTGEN} is used to generate the decays and GEANT~\cite{GEANT} is used to simulate the CLEO-c detector response.

Standard CLEO-c selection criteria are applied for $\pi^\pm$, $K^\pm$, $\pi^0$ and $K^0_{\rm S}$ candidates, as described in Ref.~\cite{CLEOCSTANDARD}.  In addition, for $K^0_{\rm S}$ candidates it is required that $|M(\pi^+\pi^-)-M(K^0_{\rm S})|<7.5~{\rm MeV}/c^2$ and the decay vertex is separated from the interaction region with a significance greater than two standard deviations.
Final states are fully reconstructed via two kinematic variables: the
beam-constrained candidate mass, $M_{bc}\equiv\sqrt{s/(4c^{4})-\mathbf{p}_{D}^{2}/c^{2}}$, where
$\mathbf{p}_{D}$ is the $D$ candidate momentum, and $\Delta E\equiv E_{D}-\sqrt{s}/2$, where $E_{D}$ is the sum
of the $D$ daughter candidate energies.   Requirements are imposed of $|\Delta E| < 20$~MeV and $|\Delta E| < 30$~MeV for the $K^-\pi^+\pi^-\pi^+$ and $K^0_{\rm S} \pi^+\pi^-$ decay modes, respectively, and $-58 < \Delta E < 35$~MeV for $K^-\pi^+\pi^0$.
The double-tagged yield is determined from counting events in the signal region of the two-dimensional $M_{bc}$ space of both meson candidates, defined by $1.86 < M_{bc} < 1.87$~GeV$/c^2$, and the sidebands used to subtract the combinatoric background.  Figure~\ref{fig:pristine} shows the data distributions for $K^{-}\pi^{+}\pi^{0}$ and $K^{-}\pi^{+}\pi^{+}\pi^{-}$ candidates tagged by $K^{0}_{\rm S}\pi^{+}\pi^{-}$; the low level of combinatoric background is clearly shown. Peaking backgrounds are estimated and subtracted using a sample of simulated $D\bar{D}$ Monte Carlo, approximately 3.3 times larger than the data.  The sizes of these  backgrounds vary from bin-to-bin, and are on average $1.3 \%$ for the $K^-\pi^+\pi^+\pi^-$ selection and $0.5\%$ for the $K^-\pi^+\pi^0$ selection. 

For double tags containing $K^-\pi^+\pi^+\pi^-$ decays, on average $1.08$ pairs of candidates are found per selected event.  In the case of $K^-\pi^+\pi^0$ decays the corresponding number is $1.15$.  In those events with more than one candidate pair, that in which the two $M_{bc}$ values most closely match the $D^0$ nominal mass is  retained for the final analysis.

A kinematic fit is applied to determine more reliably the position of a candidate in the Dalitz plot. The decay products of each $D$ candidate are constrained to have the invariant mass of the $D^0$ meson, and the $K^0_{\rm S}$-candidate daughters are constrained to the nominal  $K^0_{\rm S}$ mass.  Around $0.5\%$ of events fail this fit, or lie outside the Dalitz boundaries, and are discarded.

Simulated samples of 250,000 signal Monte Carlo events are used to determine the relative bin-to-bin efficiencies with a precision of $\sim 1\%$ (knowledge of the absolute efficiency is not important in the analysis).     This variation in efficiency is small, with the most efficient bin being a relative $13 \%$  above the lowest efficiency bin.

The yield results in each bin, $Y_i$, are given in Table~\ref{tab:yields}.  The displayed uncertainties are statistical, as the systematic biases associated with the background subtraction and efficiency correction are negligible in comparison.
Bin-to-bin migration induced by finite momentum resolution is at the level of 1 to 2\% depending upon the bin, and gives negligible bias in the coherence factor fit since the effect is  in common with that present in the analysis performed to measure the $c_i$ and $s_i$ parameters~\cite{CLEOKSPIPI}.  

\begin{table}[tb]
\caption{Bin-by-bin yields, $Y_i$, of signal decays double-tagged with $K^0_{\rm S}\pi^+\pi^-$.  The numbers are background subtracted and have been corrected for relative efficiency effects between bins, normalised to the bin of lowest efficiency.}
\label{tab:yields}
\begin{center}
\begin{tabular}{c c c c c  c}\hline\hline
Bin & $K^-\pi^+\pi^0$ & $K^-\pi^+\pi^+\pi^-$ &  Bin &
$K^-\pi^+\pi^0$& $K^-\pi^+\pi^+\pi^-$ \\ \hline
1 & $566.3 \pm 25.2$ & $370.3 \pm 20.7$ & $-1$ & $285.5 \pm 18.0$ & $194.0 \pm 15.0$ \\
2 & $267.9 \pm 17.3$  & $224.5 \pm 15.9$ & $-2$ & $67.5 \pm \hspace*{0.21cm} 8.9$    & $ 61.0 \pm \hspace*{0.21cm}  8.3$ \\
3 & $224.6 \pm 15.6$ & $190.3 \pm 14.4$ & $-3$ & $70.7 \pm  \hspace*{0.21cm} 9.0$    & $49.6 \pm \hspace*{0.21cm}  7.5$ \\
4 & $ 90.3 \pm \hspace*{0.21cm}  9.8$   & $ 60.2 \pm \hspace*{0.21cm} 8.1$   & $-4$ & $50.4 \pm  \hspace*{0.21cm} 7.2$    & $39.3 \pm \hspace*{0.21cm}  6.4$ \\
5 & $259.2 \pm 17.2$ & $185.8 \pm 14.5$ & $-5$ & $138.0 \pm 12.6$ & $99.3 \pm 10.7$ \\
6 & $210.2 \pm 15.5$ & $119.0 \pm 11.9$ & $-6$ & $ 49.9 \pm  \hspace*{0.21cm} 8.0$   & $ 34.7 \pm \hspace*{0.21cm}  6.6$ \\
7 & $403.6 \pm 21.3$ & $313.3 \pm 18.8$ & $-7$ & $44.9 \pm \hspace*{0.21cm}  7.4$    & $37.9 \pm  \hspace*{0.21cm} 6.7$ \\
8 & $426.2 \pm 21.7$ & $303.3 \pm 18.3$ & $-8$ & $93.4 \pm 10.2$   & $78.2 \pm  \hspace*{0.21cm} 9.5$ \\
\hline \hline
\end{tabular}
\end{center}
\end{table}

\section{Previously measured observables for determining the coherence factor and\\ average strong-phase difference}

\begin{table}[tb]
\caption{Measured $\rho$ and $\Delta$ observables, as reported in Ref.~\cite{WINGS}, but with modifications as explained in the text. Here the first uncertainty is statistical and the second systematic.}
\label{tab:rho_observables}
\begin{center}
\begin{tabular}{crcrcr} \hline\hline
Observable & \multicolumn{5}{c}{Measured Value} \\ \hline
$\rho^{K\pi\pi^0}_{CP+}$    & $1.119$ & $\pm$ &  $0.020$ & $\pm$ &  $0.032$ \\ 
$\rho^{K\pi\pi^0}_{CP-}$      & $0.869$ & $\pm$  & $0.023$ & $\pm$ & $0.048$ \\
$\rho^{K\pi\pi^0}_{LS}$        & $0.388$ & $\pm$  & $0.127$ &  $\pm$ & $0.026$ \\
$\rho^{K\pi\pi^0}_{K\pi,LS}$ & $0.180$ & $\pm$  & $0.076$ & $\pm$  & $0.028$ \\
$\Delta^{K\pi\pi^0}_{CP}$     &  $0.119$ & $\pm$ & $0.015$ & $\pm$ &  $0.022$ \\ \hline
$\rho^{K3\pi}_{CP+}$         &  $1.087$ & $\pm$ & $0.024$ & $\pm$ & $0.029$ \\
$\rho^{K3\pi}_{CP-}$         & $0.934$   & $\pm$ & $0.027$ & $\pm$ & $0. 046$ \\
$\rho^{K3\pi}_{LS}$          & $1.116$    & $\pm$ & $0.227$ & $\pm$ & $0.073$ \\
$\rho^{K3\pi}_{K\pi, LS}$    & $1.018$  & $\pm$ & $0.177$ & $\pm$ & $0.054$ \\ 
$\Delta^{K3\pi}_{CP}$ &  $0.084$         &$\pm$ &  $0.018$ & $\pm$ & $0.022$  \\
\hline
$\rho^{K\pi\pi^0}_{K3\pi,LS}$&    $1.218$ & $\pm$ &  $0.169$ &  $\pm$ & $0.062$   \\ \hline \hline
\end{tabular}
\end{center}
\end{table}

It is useful to recall briefly the observables measured in the original CLEO analysis.  
Double-tag yields were measured in which the signal decay, for example $K^-\pi^+\pi^0$,  is accompanied either by a decay to $CP$-even or $CP$-odd eigenstate  ($CP+$, $CP-$),  to another signal decay in which the kaon has the same charge as that of the first decay (likesign, $LS$), or by a $K^-\pi^+$ decay, again where the kaon has the same charge as that in the sister decay ($K\pi, LS$).  The observables $\rho^{K\pi\pi^0}_{CP+}$,  $\rho^{K\pi\pi^0}_{CP-}$, $\rho^{K\pi\pi^0}_{LS}$ and $\rho^{K\pi\pi^0}_{K\pi, LS}$ were then determined, which are the ratios of the measured double-tag yields to the yields expected in the absence of quantum-correlations.  The deviation of any of these observables from unity is indicative of a non-zero coherence factor.   Equivalent observables were measured for the decay  $K^-\pi^+\pi^+\pi^-$.  A ninth observable, $\rho^{K\pi\pi^0}_{K3\pi, LS}$ was defined in a similar manner for events where the double-tag is formed from the two signal decays in the case where the kaons are of the same charge.  Finally, the derived observable $\Delta^{K\pi\pi0}_{CP} \equiv \pm 1  \times (\rho^{K\pi\pi^0}_{CP \pm} - 1)$, was calculated   (and similarly  $\Delta^{K3\pi}_{CP}$), in order that the results for the $CP$-even and $CP$-odd tags could be combined together in a useful manner.   Precise definitions, and expressions relating the observables to the physics parameters of interest, can be found in Ref.~\cite{WINGS}.   

The measured values of the $\rho$ and $\Delta$ observables are summarised in Table~\ref{tab:rho_observables}.  Small corrections have been applied with respect to those values reported in Ref.~\cite{WINGS} to take account of the improved knowledge of the $D^0 \bar{D}{}^0$ mixing parameters \cite{HFAG}, the CF $D^0 \to K^-\pi^+\pi^0$ and  $D^0 \to K^-\pi^+\pi^+\pi^-$ branching ratios~\cite{NEWCLEO}, and the DCS $D^0 \to K^+\pi^-\pi^+\pi^-$ branching ratios~\cite{ERIC}, all of which enter the analysis.
It can be seen that the deviations from the zero-coherence hypothesis are generally more significant for the $K^-\pi^+\pi^0$  observables than for those of $K^-\pi^+\pi^+\pi^-$.

\section{Fit results}

\begin{table}[ht]
 \caption{Values of mixing parameters and branching fractions used in the fit.}\label{tab:inputs}
\begin{center}
\begin{tabular}{lcc} \hline\hline
Parameter & Value & Reference \\ \hline
 $x$ & $(0.39^{+0.17}_{-0.17})\%$ & \cite{HFAG} \\
 $y$ & $(0.67^{+0.07}_{-0.08})\%$ & \cite{HFAG} \\
 $\delta_{D}^{K\pi}$ & $(192.5^{+\phantom{0}9.4}_{-11.0})^{\circ}$ & \cite{HFAG} \\ 
$\mathcal{B}(D^{0}\to K^{-}\pi^{+}\pi^{0})$ & $(14.96\pm 0.34)\%$ & \cite{NEWCLEO}  \\
$\mathcal{B}(D^{0}\to K^{+}\pi^{-}\pi^{0})$ & $(3.28\pm 0.20)\times 10^{-4}$ & \cite{PDG,NEWCLEO}  \\
$\mathcal{B}(D^{0}\to K^{-}\pi^{+}\pi^{+}\pi^{-})$ & $(8.29\pm 0.20)\%$ & \cite{NEWCLEO}  \\
$\mathcal{B}(D^{0}\to K^{+}\pi^{-}\pi^{-}\pi^{+})$ & $(2.68\pm 0.11)\times 10^{-4}$ & \cite{ERIC,NEWCLEO}  \\
$\mathcal{B}(D^{0}\to K^{-}\pi^{+})$ & $(3.88\pm 0.05)\%$ & \cite{PDG}  \\
$(r_{D}^{K\pi})^2$ & $(0.349 \pm 0.004) \%$ & \cite{HFAG}  \\ \hline\hline
\end{tabular}
\end{center}
\end{table}

\begin{figure}[ht]
\begin{center}
\begin{tabular}{cc}
\includegraphics[width=0.45\columnwidth]{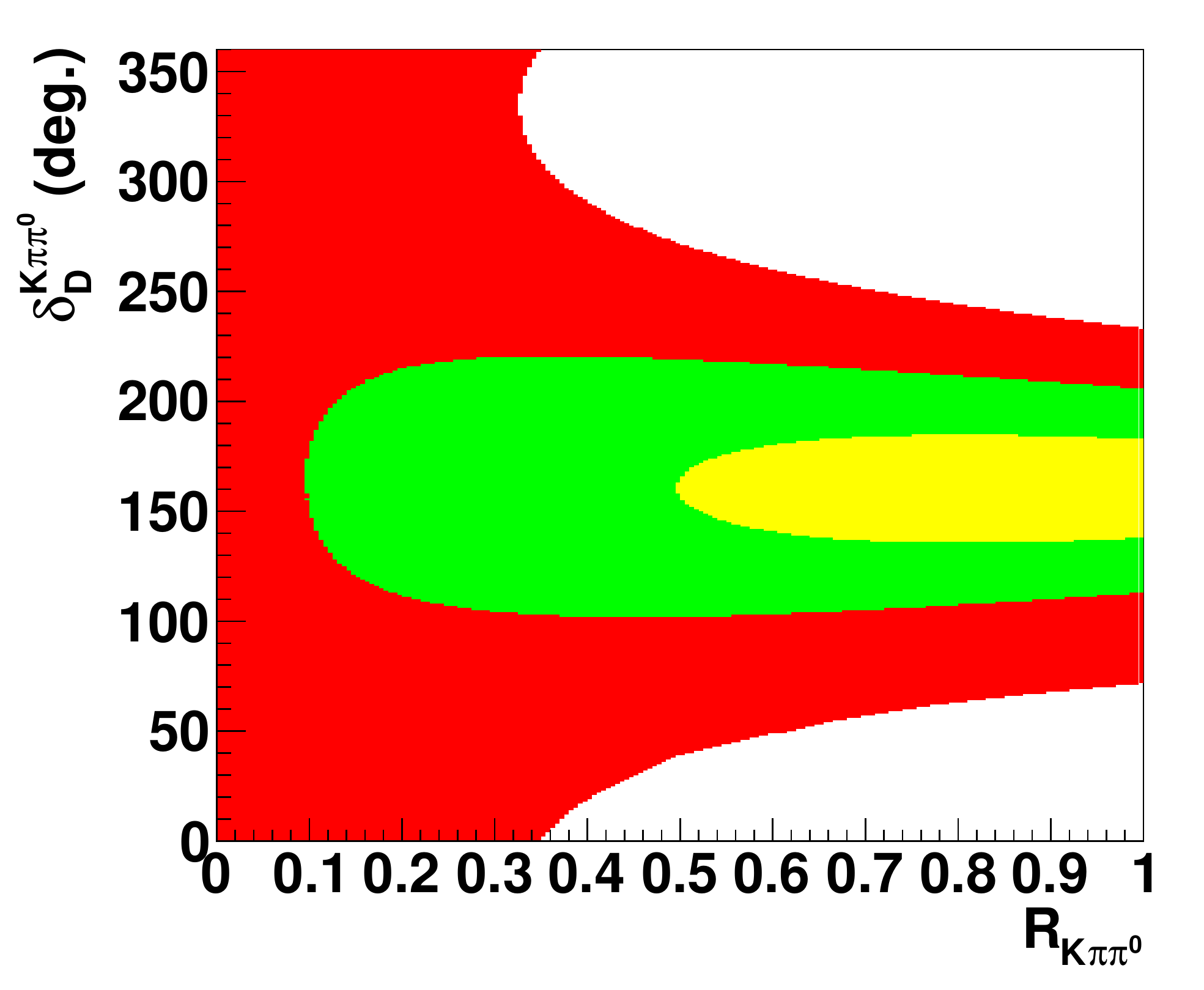}  &
\includegraphics[width=0.45\columnwidth]{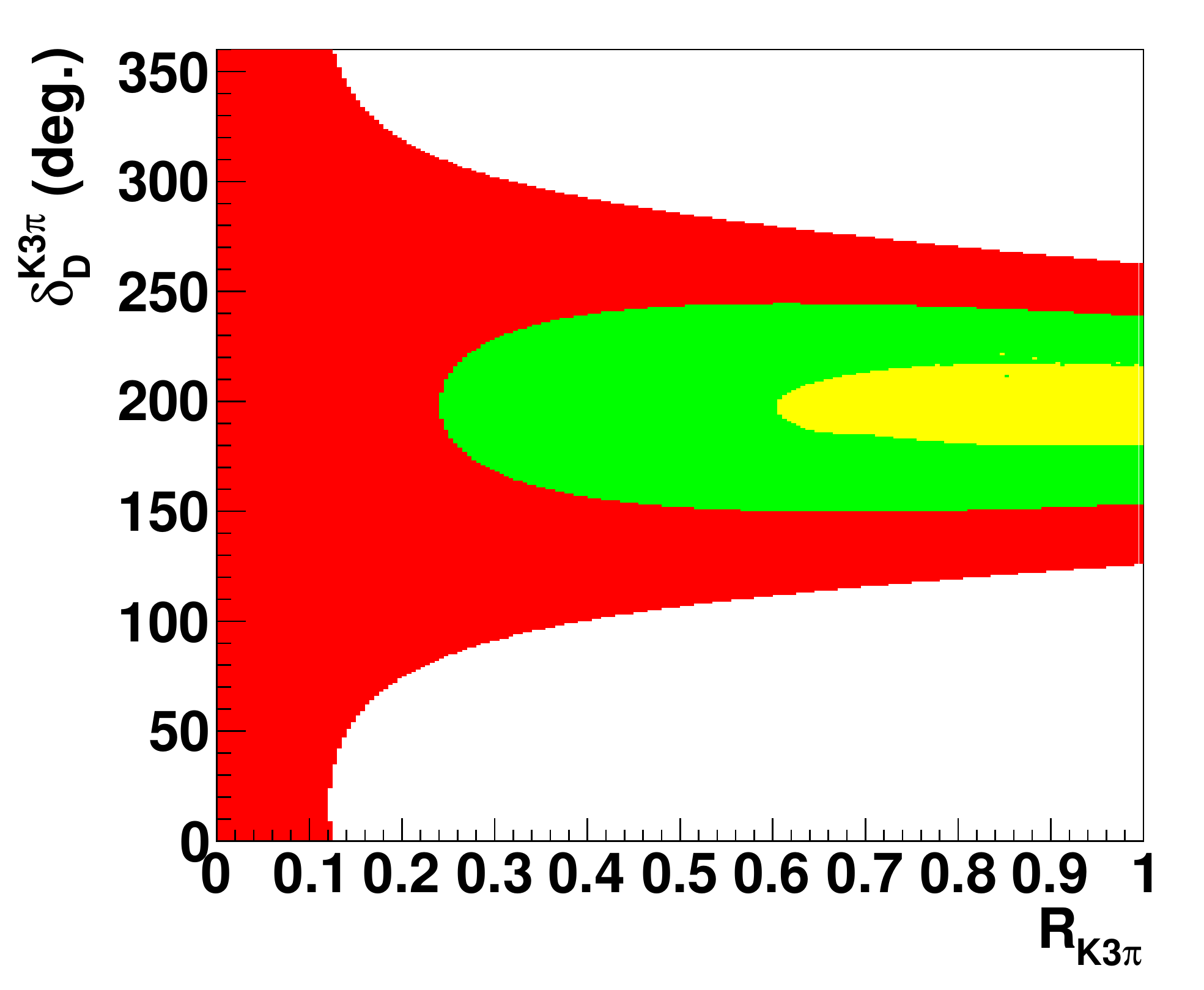}  \\
\end{tabular}
\caption{[Colour online] Scans of the contribution from the $Y_i$ observables to the $\Delta\chi^2$ in the (left) $(R_{K\pi\pi^{0}},\delta_{D}^{K\pi\pi^{0}})$ and (right) $(R_{K3\pi},\delta_{D}^{K3\pi})$ parameter space. The colours correspond to $1\sigma$ (yellow), $2\sigma$ (green) and $3\sigma$ (red) confidence intervals.}\label{fig:pseudoscans}
\end{center}
\end{figure}

\begin{figure}[ht]
\begin{center} 
 \begin{tabular}{cc}
\includegraphics[width=0.45\columnwidth]{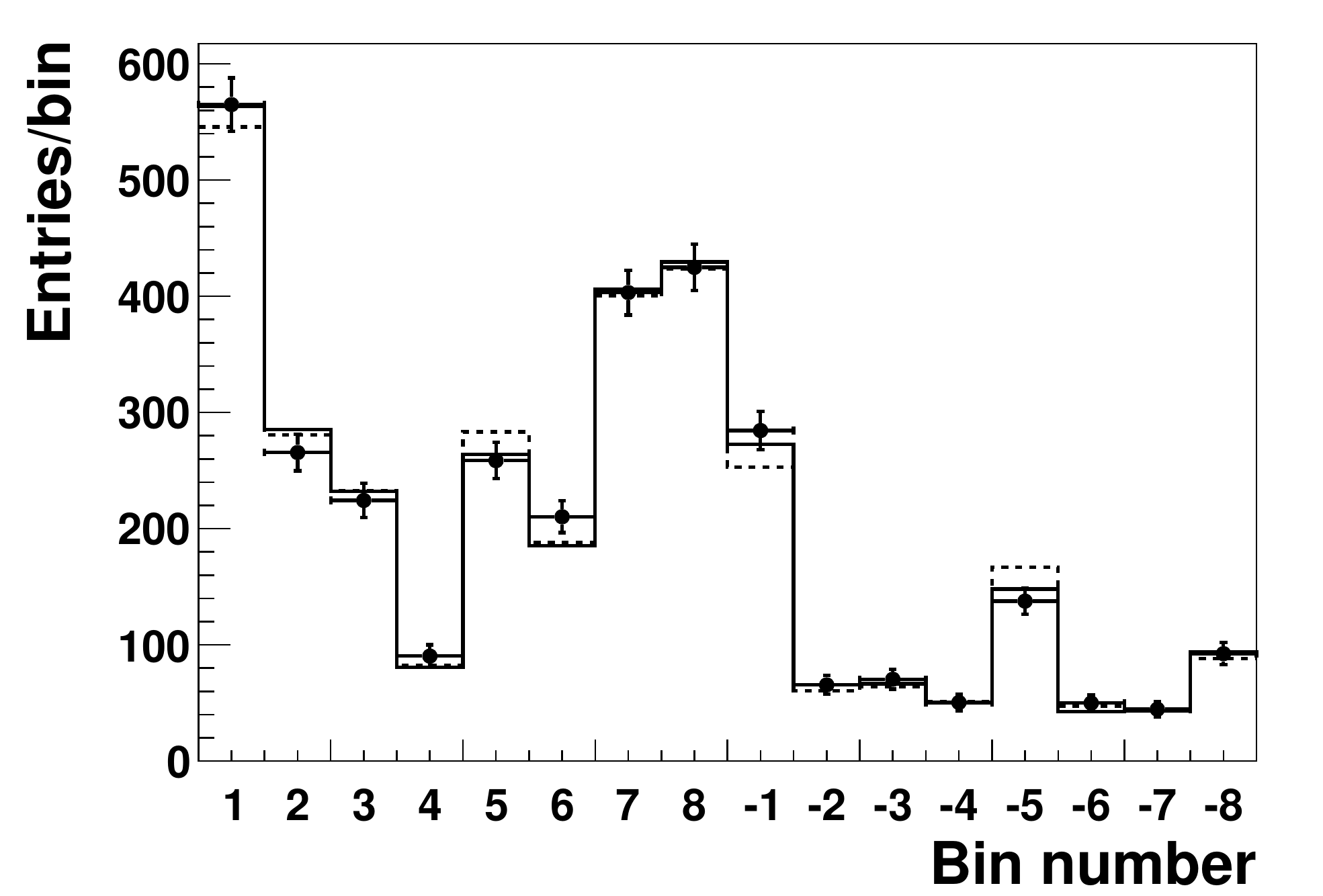} & \includegraphics[width=0.45\columnwidth]{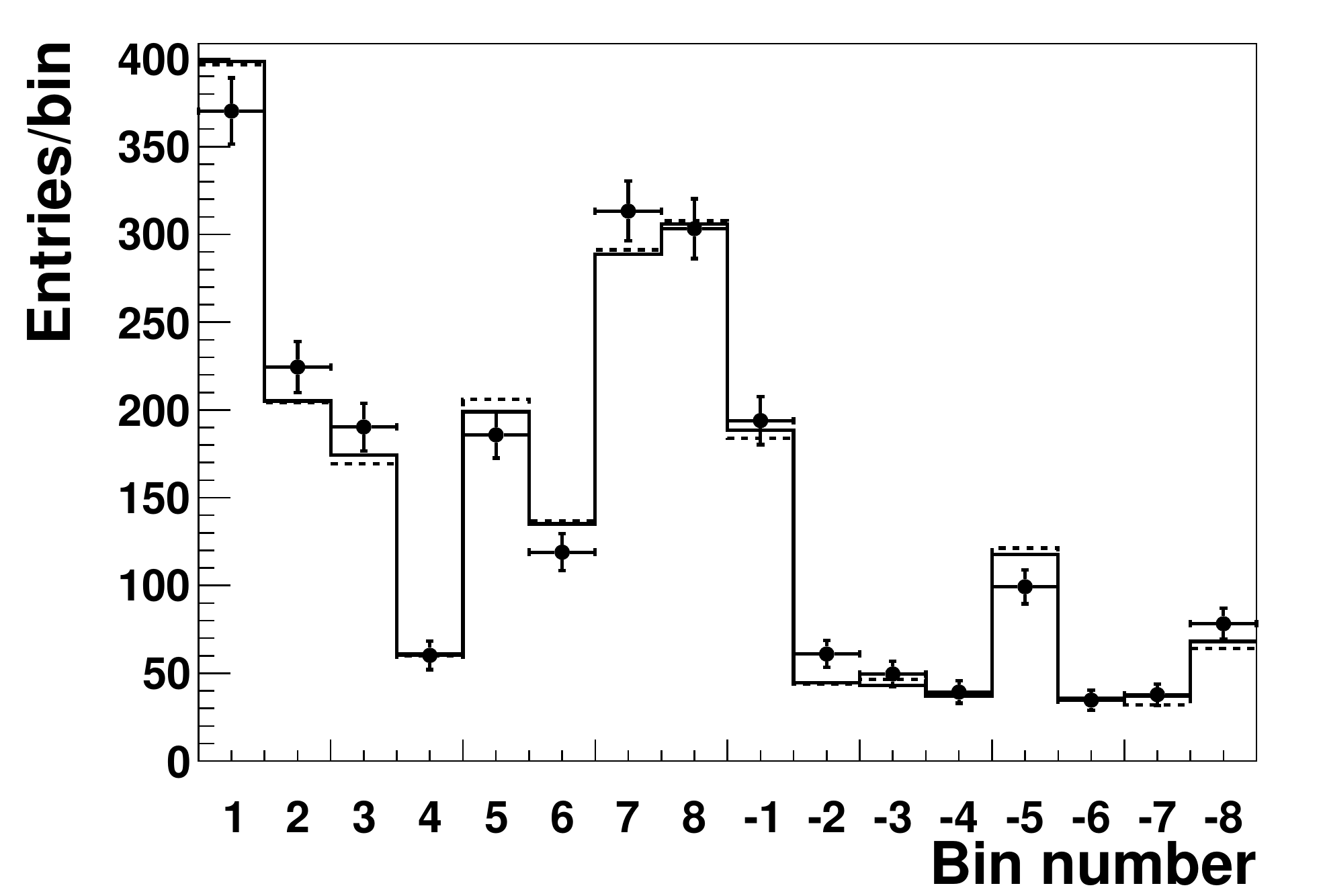}
 \end{tabular}
\end{center}
\caption{Comparison of the (left) $K^{-}\pi^{+}\pi^{0}$ vs. $K^{0}_{\rm S}\pi^{+}\pi^{-}$ $Y_i$ and (right) $K^{-}\pi^{+}\pi^{+}\pi^{-}$ vs. $K^{0}_{\rm S}\pi^{+}\pi^{-}$ data (points with error bars) with the expectation from the best fit values of the parameters (solid line). Also shown is the expected values of $Y_i$ if there was no coherence in the decay (dashed line).}\label{fig:pseudobinbybin}
\end{figure}

All measurements of the observables, along with the corresponding covariance matrix, are combined in a $\chi^2$ fit to determine $R_{K\pi\pi^{0}}$, $\delta_{D}^{K\pi\pi^{0}}$, $R_{K3\pi}$ and $\delta_{D}^{K3\pi}$. In addition, the other parameters on which the observables depend - $x$, $y$, $\delta_{D}^{K\pi}$, $r_D^{K\pi}$, $c_i$, $s_i$, $K_i$ and the $D^{0}$  branching fractions  - are free parameters in the fit. However, these external parameters are Gaussian constrained to their measured central values. The values and uncertainties used for $x$, $y$, $\delta_D^{K\pi}$, $r_{D}^{K\pi}$ and the branching fractions are given in Table~\ref{tab:inputs}; apart from $\mathcal{B}(D^{0}\to K^{-}\pi^{+})$ all external parameters have been updated since the original CLEO analysis. In particular, the CLEO collaboration has reported new values of the CF branching fractions for $D^{0}\to K^{-}\pi^{+}\pi^{0}$ and $D^{0}\to K^{-}\pi^{+}\pi^{+}\pi^{-}$ decays~\cite{NEWCLEO}; these values are used to normalise the measurements of the DCS branching fractions reported in Refs.~\cite{PDG,ERIC}. Therefore, the DCS branching fractions used in the fit have been scaled appropriately to reflect this change in the value of the normalising branching fraction.  The values of $c_i$ and $s_i$, along with their uncertainties and correlations, are taken from Ref. \cite{CLEOKSPIPI}. The values of $K_i$ are those given in Table~\ref{tab:kip}.

The additional information the $Y_i$ observables bring to the analysis can be seen in Fig.~\ref{fig:pseudoscans} where scans of the $\Delta\chi^2$ for the new observables alone are shown over the $(R_{K\pi\pi^0},\delta_D^{K\pi\pi^0})$ and $(R_{K3\pi},\delta_D^{K3\pi})$ parameter spaces. The scans indicate that significant coherence in both modes is favoured by the $Y_i$ observables. Furthermore, the expected values of the $Y_i$ observables for the best fit values of the coherence parameters is 
compared to the data for the $K^{-}\pi^{+}\pi^{0}$ vs. $K^{0}_{\rm S}\pi^{+}\pi^{-}$ and $K^{-}\pi^{+}\pi^{+}\pi^{-}$ vs. $K^{0}_{\rm S}\pi^{+}\pi^{-}$ data in Fig.~\ref{fig:pseudobinbybin}. Also, shown are the values of $Y_i$ expected if there was no coherence to indicate the variation of these observables with significant coherence.

\begin{figure}[ht]
\begin{center}
\includegraphics[width=0.6\columnwidth]{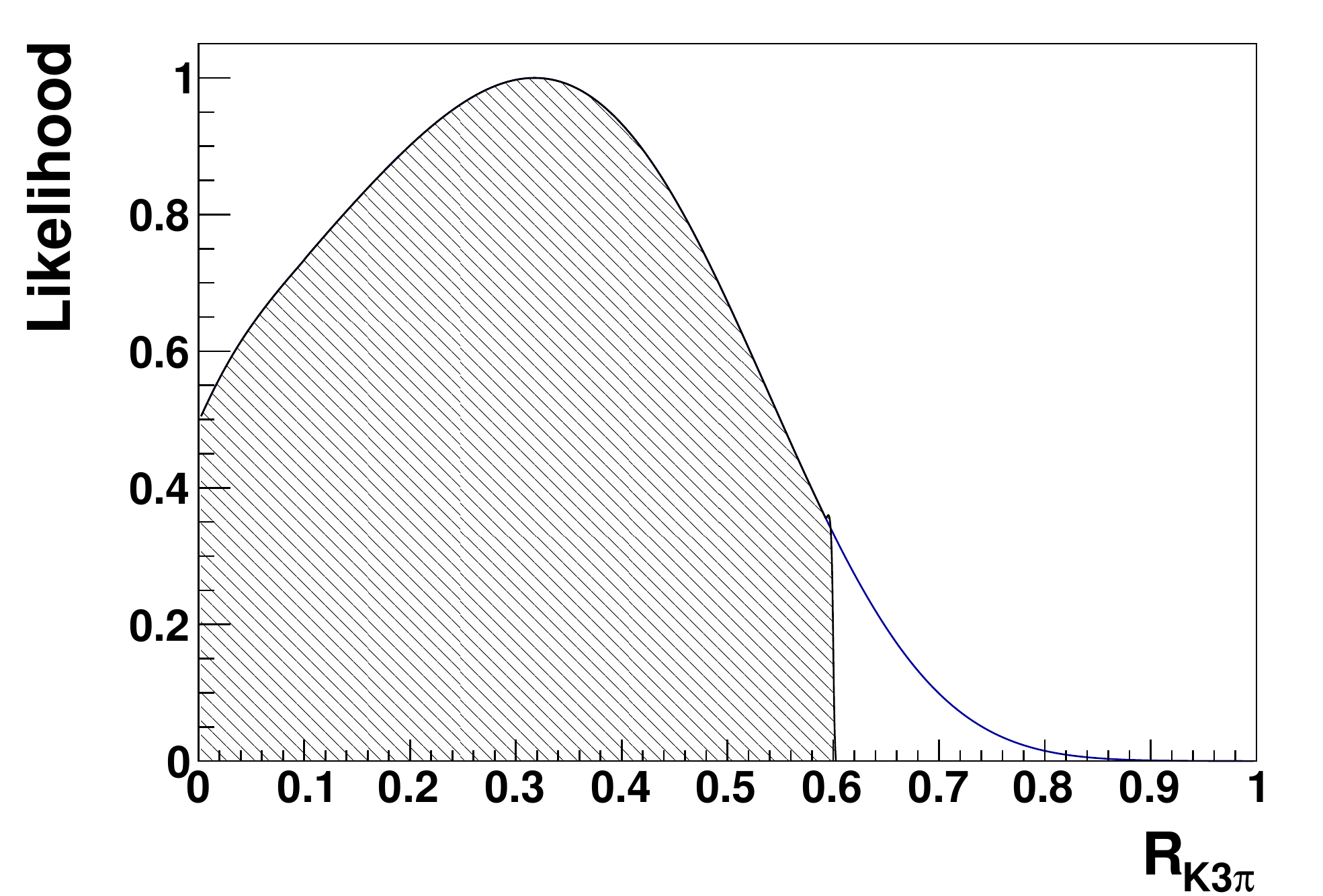}
\end{center}
\caption{Likelihood scan of $R_{K3\pi}$ within the physical region. The 95\% confidence level interval is indicated by the hatched region.}\label{fig:R_K3piscan}
\end{figure}

The best fit values and the correlations amongst the parameters  for the global fit are given in Tables~\ref{tab:results} and \ref{tab:correl}, respectively. The reduced $\chi^2$ of the fit is 44.4/33, which corresponds to a probability of 8.9\%.  The best fit value of $R_{K3\pi}$ is compatible with zero within two standard deviations. Therefore, a scan of the likelihood within the physical region $R_{K3\pi}\in \{0,1\}$ is performed to determine a confidence interval for this parameter. The likelihood is 
\begin{displaymath} 
\mathcal{L}(R_{K3\pi}) = \mathrm{exp}\left[-\: \frac{\chi^2_{0}-\chi^2(R_{K3\pi})}{2}\right] \; ,
\end{displaymath} 
where $\chi^2_0$ is the best fit value of the $\chi^2$ with all parameters free and  $\chi^2(R_{K3\pi})$ is the best fit value of the $\chi^2$ with $R_{K3\pi}$ fixed but all other parameters free. The resulting likelihood scan is shown in Fig.~\ref{fig:R_K3piscan}. The upper 95\% confidence limit $R_{K3\pi}^{95\%}$ is defined as
\begin{displaymath}
 \frac{\int_{0}^{R_{K3\pi}^{95\%}} \mathcal{L}(R_{K3\pi})\: \mathrm{d}R_{K3\pi}}{\int_{0}^{1} \mathcal{L}(R_{K3\pi})\: \mathrm{d}R_{K3\pi}} = 0.95 \; .
\end{displaymath}    
The resulting upper 95\%  confidence limit is $R_{K3\pi}< 0.60$. 

\begin{table}[ht]
\caption{Results from the fit. The uncertainties are the combination of the statistical and systematic uncertainties.}\label{tab:results}
\begin{center}
\begin{tabular}{lc} \hline\hline
Parameter & Fitted value \\ \hline
$R_{K\pi\pi^{0}}$             &  $0.82\pm 0.07$                \\ 
$\delta_D^{K\pi\pi^{0}}$ &  $(164^{+20}_{-14})^{\circ}$       \\ 
$R_{K3\pi}$                       &  $0.32^{+0.20}_{-0.28}$ \\
$\delta_D^{K3\pi}$           &  $(255^{+21}_{-78})^{\circ}$ \\ \hline\hline 
\end{tabular}
\end{center}
\end{table}

\begin{table}[ht]
   \caption{Correlation coefficients between the parameters.}\label{tab:correl} 
\begin{center}
\begin{tabular}{l c c c} \hline\hline
                                              &  $\delta_D^{K\pi\pi^{0}}$ & $R_{K3\pi}$ &  $\delta_{D}^{K3\pi}$ \\ \hline
$R_{K\pi\pi^{0}}$                &  $-0.444$                             & $\phantom{-}0.216$       &  $-0.008$   \\
$\delta_D^{K\pi\pi^{0}}$    &     --                                      &  $-0.477$      &  $\phantom{-}0.097$  \\
 $R_{K3\pi}$                         &    --                                       &  --                  &  $\phantom{-}0.201$ \\ \hline\hline
\end{tabular} 
 \end{center}
\end{table}

 In the original CLEO analysis \cite{WINGS} some improvement was observed in the $y$ and $\delta_D^{K\pi}$ uncertainties with these parameters constrained in the fit; this motivated a fit with these parameters unconstrained to determine the standalone sensitivity to the charm-mixing parameters. However, given the improvements in the determination of the charm-mixing parameters \cite{HFAG} since the original CLEO analysis, such a reduction in uncertainty  is no longer observed. Therefore, a fit with the charm-mixing parameters unconstrained is not presented here. 

\begin{figure}
\begin{center}
\begin{tabular}{cc}
\includegraphics[width=0.45\columnwidth]{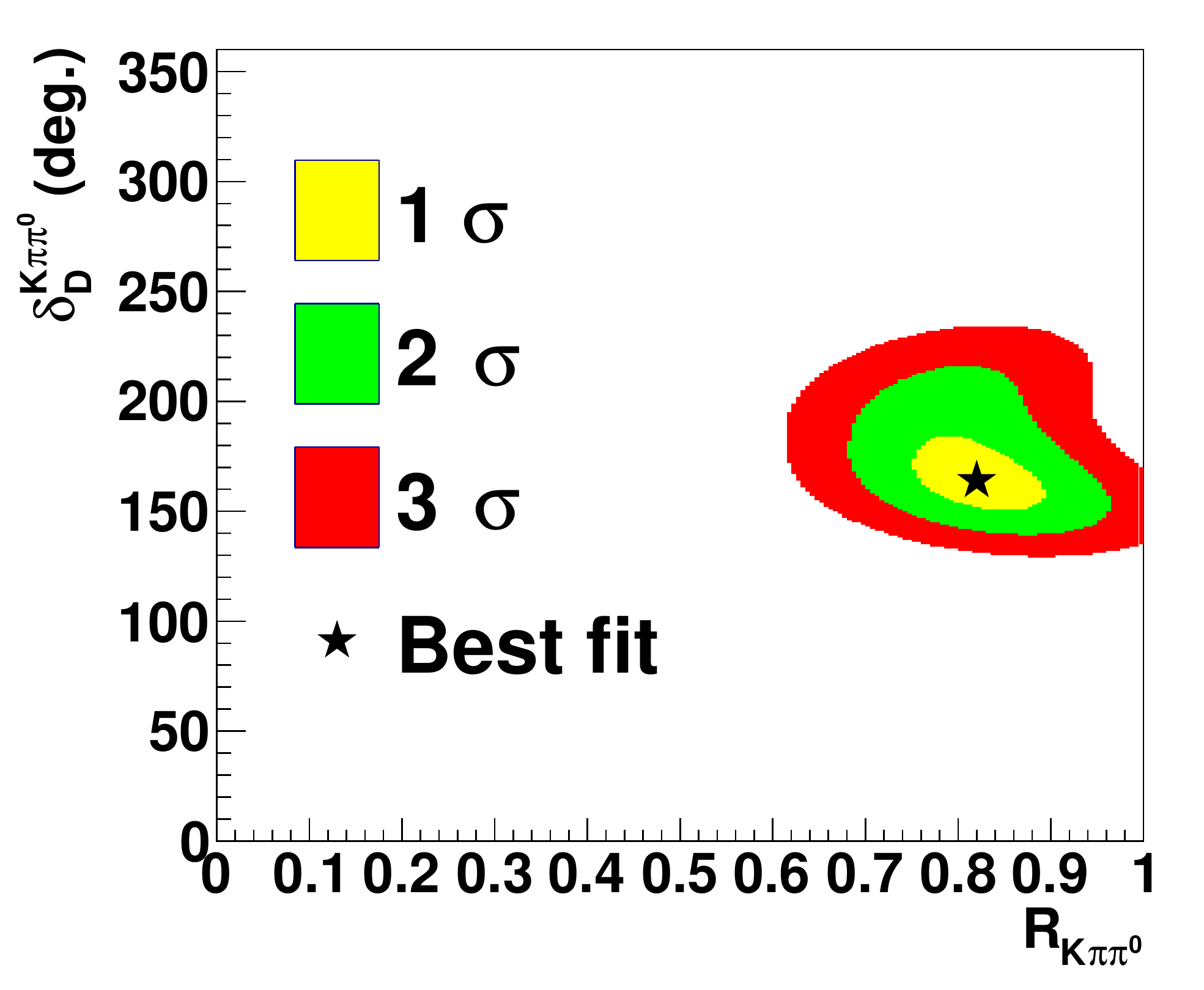}  &
\includegraphics[width=0.45\columnwidth]{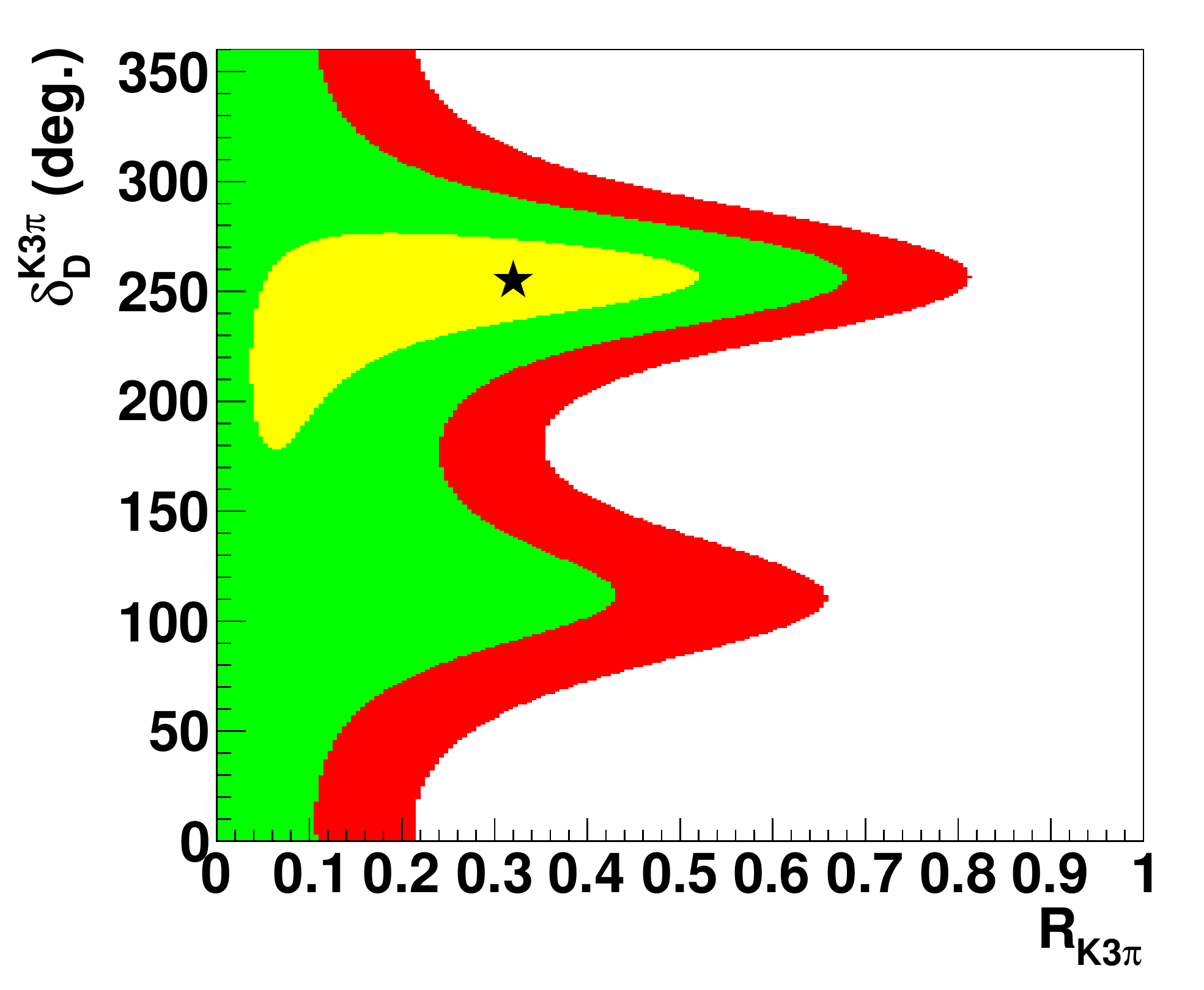}  \\
\end{tabular}
\caption{[Colour online] Scans of the $\Delta\chi^2$ in the (left) $(R_{K\pi\pi^{0}},\delta_{D}^{K\pi\pi^{0}})$ and (right) $(R_{K3\pi},\delta_{D}^{K3\pi})$ parameter space.}\label{fig:scans}
\end{center}
\end{figure}

Scans of the $(R_{K\pi\pi^{0}},\delta_{D}^{K\pi\pi^{0}})$ and $(R_{K3\pi},\delta_{D}^{K3\pi})$ parameter space are shown in Fig.~\ref{fig:scans}. The $\Delta\chi^2$ is used to determine the one, two and three standard deviation confidence intervals within the parameter space. The values of $R$ and $\delta$ are fixed, while all other parameters are left free while minimising the $\chi^2$ at each point from which a $\Delta\chi^2$ with respect to $\chi^2_{0}$ can be obtained. These scans indicate the non-Gaussian nature of the confidence regions for $R$ and $\delta_{D}$. Therefore, when these results are to be used in an analysis it is recommended to use the full $\Delta\chi^2$ scan \cite{SCANS} or the observables themselves.

\section{Outlook and conclusions}
Updated measurements of the coherence factors and average strong-phase differences for $D^{0}\to K^{-}\pi^{+}\pi^{0}$ and $D^{0}\to K^{-}\pi^{+}\pi^{+}\pi^{-}$ have been presented. Despite the addition of events tagged by $D^{0}\to K^{0}_{\rm S}\pi^{+}\pi^{+}$ decays the overall precision on the parameters has not improved significantly compared to the original CLEO-c analysis \cite{WINGS}. However, the likelihood curves are significantly different to those previously published as a result of the changes in the central values of the parameters, in particular those of the average strong-phase differences.
These changes are due to the additional data and the updates to the $D^{0}$ branching fractions and charm-mixing parameters. Therefore, it is recommended that the new results are used in the determination of  $\gamma/\phi_3$ from $B^{\pm}\to DK$ decays and in charm-mixing studies. 
  
 The BESIII detector \cite{BESIII} has collected a correlated $D\bar{{D}}$ data set at a centre-of-mass energy corresponding to the mass of the $\psi(3770)$.  This data set is approximately 3.5 times larger than that used in this analysis. 
An estimate of the BESIII potential to determine the coherence factors and strong-phase differences is obtained by reducing the uncertainties on the observables and $Y_i$ measurements by a factor of $1/\sqrt{3.5}$, then repeating the $\chi^2$ fit to the parameters. The uncertainties returned by the fit are: $\sigma(R_{K\pi\pi^{0}})= 0.04$, $\sigma(\delta^{K\pi\pi^{0}}_D)= 8^{\circ}$, $\sigma(R_{K3\pi})= 0.10$, and $\sigma(\delta^{K3\pi}_D)= 8^{\circ}$. The uncertainties are not only reduced but symmetric. Therefore, it is clear that significant improvements in the knowledge of these parameters can be obtained from the current BESIII data set.


\section*{Acknowledgments}

This analysis was performed using CLEO-c data, and as members of the former CLEO collaboration we thank it for this privilege.
We also acknowledge useful discussions with former CLEO colleagues and Anton Poluektov.  We are grateful for support from 
the UK Science and Technology Facilities Council and the UK India and Education Research Initiative.





\end{document}